\documentclass[prl,twocolumn,preprintnumbers,amsmath,amssymb]{revtex4}
\usepackage{graphicx}
\usepackage{color}
\usepackage{url, hyperref}
\usepackage{amssymb, dsfont}

\begin{document}

\title{From 3D to 2D Hydrodynamics in Interacting Micro-rods}

\author{
R. Di Leonardo$^{1,2}$, 
E. Cammarota$^2$,
G. Bolognesi$^3$
}

\affiliation{
$^1$IPCF-CNR, UOS Roma, P.le A. Moro 2, I-00185, Roma, Italy.\\
$^2$Dipartimento di Fisica, Universit\`a di Roma ``Sapienza'', I-00185, Roma, Italy\\
$^3$ Dipartimento di Ingegneria Meccanica e Aerospaziale, Universit\`a di Roma ``Sapienza", I-00184, Roma, Italy.
}

\begin{abstract}

Moving micron scale objects are strongly coupled to each other by hydrodynamic
interactions. The strength of this coupling decays as the inverse particle
separation when the two objects are sufficiently far apart. It has been
recently demonstrated that the reduced dimensionality of thin fluid layer gives
rise to longer ranged, logarithmic coupling. Using holographic tweezers we show
that microrods display both behaviors interacting like point particle in 3D at
large distance and like point particles in 2D for distances shorter then their
length.  We derive a simple analytical expression that fits remarkably well our
data and further validate it with finite element analysis.


\end{abstract}

\maketitle

Flagella, microtubules, nanotubes and nanowires are a few examples of the rich
variety of fundamental roles that slender bodies have in physics, chemistry and
biology. They all share the same geometrical feature of being ``slender'', that
is having a linear dimension which is much larger than the other two. As a
consequence of that they're all expected to have a similar physical behavior in
those situations where only shape matters. That is, for example, the case for
micro-hydrodynamics. A thin nanotube or a thicker microrod will both experience
a similar drag force, which will depend mainly on the body length while
thickness only appears in logarithmic terms \cite{happel}. While there's a huge
amount of work on single slender body dynamics, especially in the context of
bacterial motility \cite{lighthill}, investigation of hydrodynamic interactions
between slender bodies is almost only limited to the phenomenology of
syncronization \cite{reich, polin}.  One reason could be that, as opposed to
spheres, which have been studied extensively \cite{crocker, meiners, ring},
hydrodynamic couplings between anisotropic bodies are a complex function of
both relative distance and orientation.  Optical tweezers can be used to trap and
move one dimensional objects \cite{tan, onofrio, marago}. In particular,
holographic optical trapping has been shown to be an ideal tool for full 3D
micromanipulation of microrods and nanotubes \cite{plewa, grier, simpson, ikin,
carberry}. Such capabilities offer a unique opportunity for trapping and
orienting slender bodies in well defined relative configuration, and directly
probe their coupled Brownian dynamics.

In this Letter we provide a direct measurement of hydrodynamic coupling between
a pair of parallel aligned silica microrods optically trapped in blinking
holographic tweezers.  We found a crossover from 3D behavior at large
distances, to a 2D logarithmic behavior, when the distance falls below the rods
length. Experimental data are in excellent agreement with finite element
analysis and can be very well reproduced by a simplified theoretical
approach.

\begin{figure}[ht]
\includegraphics[width=.45\textwidth]{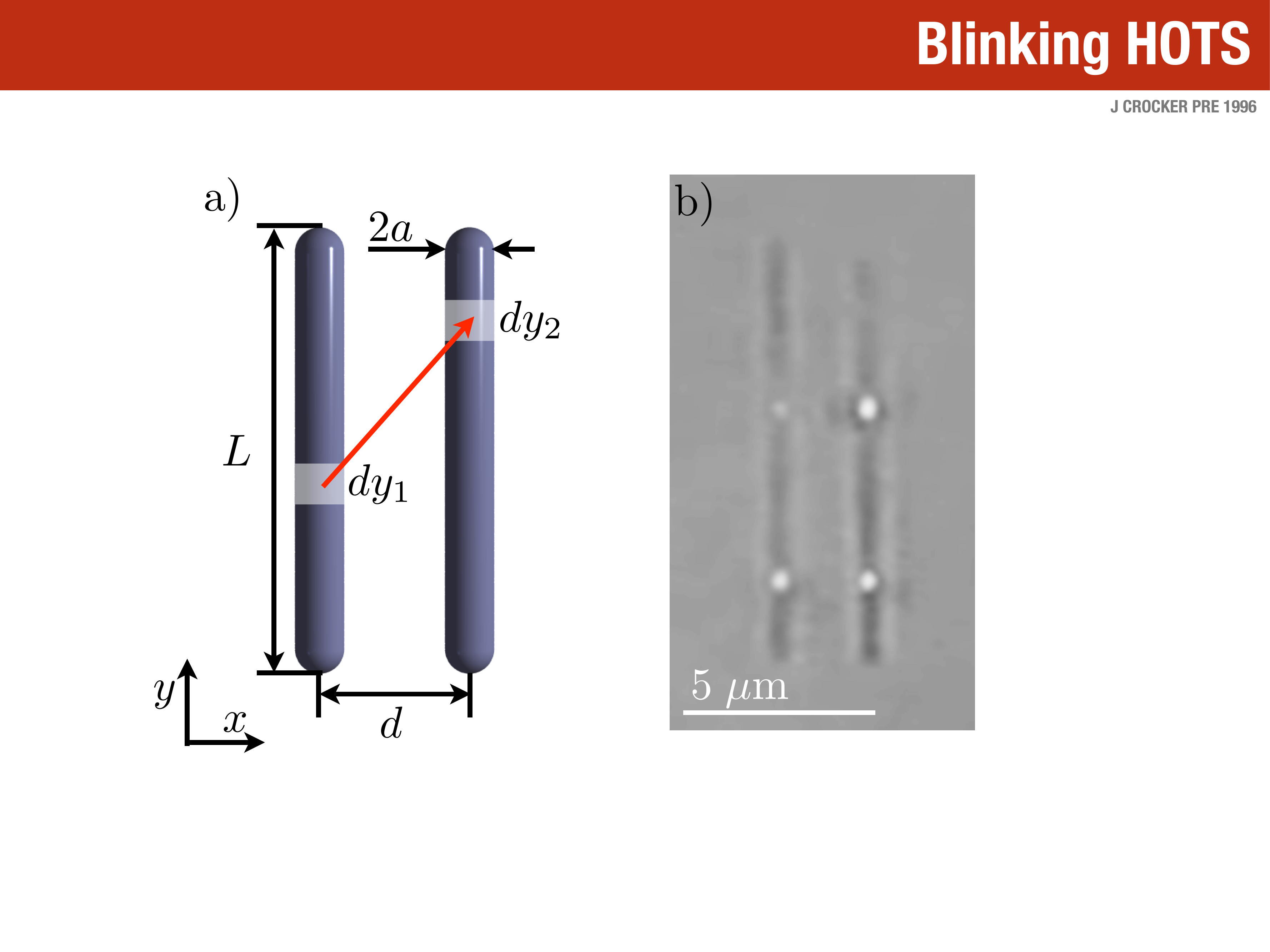}
\caption{a) Geometrical parameters of the investigated microrods configurations.
b) Two silica microrods, each one held in a double holographic trap.}
\label{rods}
\end{figure}

In a practically zero Reynolds number regime,
as it is in the mesoscopic world, a moving objects generates a perturbation in
the surrounding fluid that decays as the inverse distance \cite{happel, Kim}. A
nearby particle will be transported by this flow and experience a higher
mobility when trying to move along the same direction of the other particle.
For the same reasons relative motions are characterized by lower mobilities.
For well separated objects, collective and relative mobilities can be easily
derived assuming that each object is rigidly transported by the approximately
uniform flow field produced by the other one.  Within this approximation when
forces $\mathbf F_1$ and $\mathbf F_2$ are applied to two bodies they will move
with  speeds:
\begin{equation} \begin{array}{rcl} \mathbf U_1&=&\mathbf M_1\cdot \mathbf F_1
+ \mathbf u_{12}\\ \mathbf U_2&=&\mathbf M_2\cdot \mathbf F_2 + \mathbf u_{21}
\end{array} \end{equation}
where $\mathbf M_i$ is the ith body mobility tensor while $\mathbf u_{ij}$
represents the flow that object $j$ produces at the location of $i$. Due to
linearity in Stokes flow, flow fields have the form:
\begin{equation} \mathbf u_{ij}=\mathbf G_{ij}\cdot\mathbf F_j \end{equation}
the flow propagator $\mathbf G_{ij}$ will depend on distance and on the
relative orientation of particle $j$.  Introducing the velocity and force
vectors $\mathds U=\left( \mathbf U_1, \mathbf U_2, \dots\right)$ and $\mathds
F=\left( \mathbf F_1, \mathbf F_2, \dots\right)$, the full many-body problem
can be then stated in the form of a compact mobility matrix formulation:
\begin{equation} \mathds U=\mathds M\cdot \mathds F \end{equation}
where $\mathds M$ is the mobility matrix:
\begin{equation} \mathds M=\left( \begin{array}{ccc} \mathbf M_1& \mathbf
G_{12}&\dots\\ \mathbf G_{21}& \mathbf M_2&\dots\\ \vdots&\vdots&\ddots\\
\end{array} \right) \end{equation}
with hydrodynamic couplings appearing as off-diagonal terms.  At large enough
distances, whatever is the shape of $j$, the propagator will tend to to the
Oseen tensor that only depends on the position vector $\mathbf r_{ij}=\mathbf
r_i-\mathbf r_j$:
\begin{equation} \mathbf G_O(\mathbf r_{ij})= \frac{1}{8\pi\mu\;r_{ij}}
\left(\mathds 1 + \hat{\mathbf r}_{ij} \hat{\mathbf r}_{ij}\right)
\end{equation}
The Oseen tensor represents the first term in the multipole expansion of the
Stokes flow produced by a given force spatial distribution, also known as the
Stokeslet. The Stokeslet propagates the perturbation of a zero-dimensional
point force and therefore does not depend on particle shape and orientation.
For spherical beads  it provides a remarkably good description of couplings
down to interparticle distances of about 2.5 radii \cite{crocker, meiners,
ring}.  For anisotropic bodies, higher order terms will have to be included and
their form will depend on particle shape and orientation. A thin rod may be
thought as the simplest anisotropic body consisting of a one dimensional line force
distribution.  Most of the shape dependence occurs via a single parameter, that
is the body length $L$, and rod orientation matters. Such a dependence on
orientations, coupled to the long range character of Stokes propagators, makes
hydrodynamic coupling between thin rods a very challenging problem both from
the experimental and theoretical point of view. As a first starting approach
we could limit ourselves to a single pair of interacting rods at a fixed
distance. Slender body theory provides a practical theoretical framework for
unidimensional objects where the point force description of the sphere is
substituted by a line distribution of point forces.  Slender body theory has
been successfully applied in the context of flagellar and ciliar propulsion in
bacteria. 
We have seen that coupling terms
in the mobility matrix can be thought, as a first approximation, as the flow
propagators between interacting objects. Although this is straightforward in
the case of spheres, where sphere's center is a good representative point of
sphere position and center of forces, it is quite ambiguous when bodies of
extended size are considered.  It can be shown \cite{Kim} that when a thin
prolate spheroid is immersed in an external inhomogeneous flow, it will move
with a speed given by the average of the original flow over the rod's length.
This finding suggests that we can approximate the hydrodynamic coupling between
the rods as the average flow produced by one rod over the length of the other
one.  For parallel aligned rods like in Fig. \ref{rods} this amounts to the
double Stokeslet integration:
\begin{widetext}
\begin{equation}
\label{Gth}
G^{xx}(d)=\int_{-L/2}^{L/2} \!\!\! dy_2 \int_{-L/2}^{L/2} \!\! dy_1 
\frac{1}{8\pi\mu}\left(
\frac{1}{\sqrt{d^2+(y_2-y_1)^2}}
+\frac{d^2}{d^2+(y_2-y_1)^2}
\right)=
\frac{1}{8\pi \mu L}\log\left[\frac{L+\sqrt{L^2+d^2}}{\sqrt{2L^2+d^2-2L\sqrt{L^2+d^2}}}\right]
\end{equation}
\end{widetext}

\begin{figure}[htb]
\includegraphics[width=.48\textwidth]{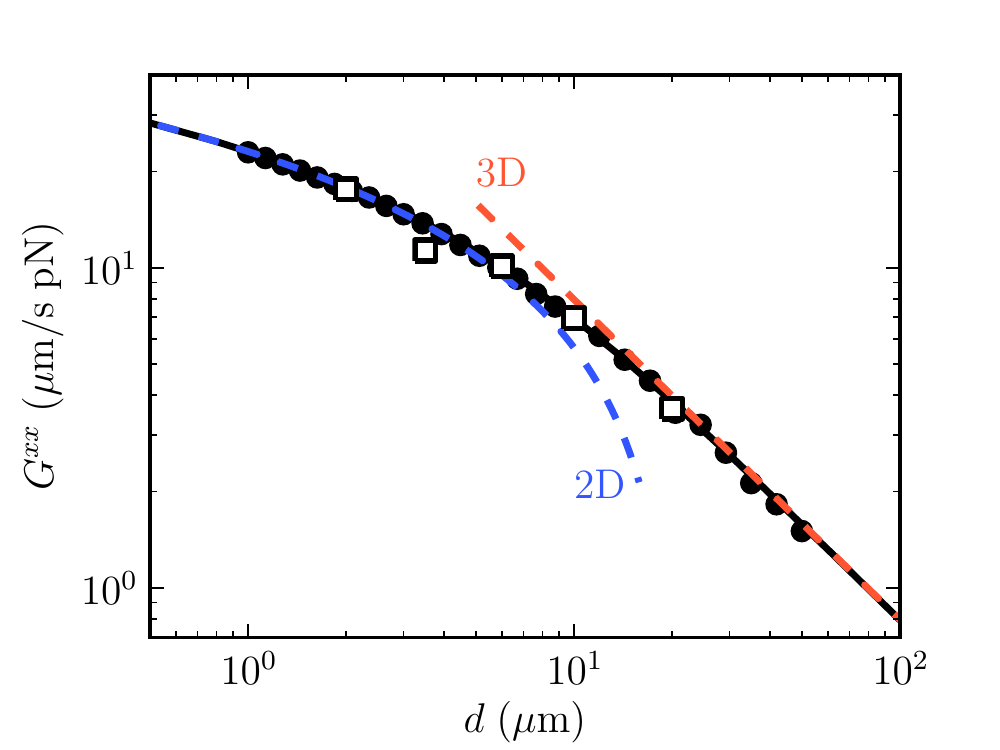}
\caption{ Hydrodynamic coupling $G^{xx}$ between two parallel aligned thin rods
as a function of their relative distance. Solid line represents the simple
theory in eq. (\ref{Gth}). Black circles are numerical results from finite
element analysis. Experimental data are reported as open squares. Dashed lines
represent the long (\ref{Gthl}) and short (\ref{Gths}) distance limiting
behavior reducing to respectively 3D and 2D hydrodynamic couplings between
point particles.}
\label{G}
\end{figure}

When the two rods are much far apart than their length, the obtained propagator
reduces obviously to the Oseen tensor component:
\begin{equation}
\label{Gthl}
G^{xx}(d\gg L)\simeq\frac{1}{4\pi\mu d}. 
\end{equation}
More interestingly, for separations that are much smaller than rods length, the
propagator has the logarithmic expression: 
\begin{equation}
\label{Gths}
G^{xx}(d\ll L)\simeq \frac{1}{4\pi\mu L} \log \frac{2 L}{d}
\end{equation}
Such a form reminds very closely of what is observed for hydrodynamic coupling
in two dimensional thin liquid layers \cite{2d}. In other words, microrods
interact like point particles in 3D when they are very far away, and then like
point particles in 2D at short distances.  The crossover distance is determined
by the most important length in the problem, that is the rod's length. Such a
finding is not surprising when one realizes that at short distances the problem
reduces to the idealized two dimensional case of infinitely long cylinders,
where the finite length of the two rods is only responsible for edge effects.
Our simple formula seems to provide a good interpolation between the two
limiting cases but too crude approximations might be involved in its derivation
and we decided to test it against finite element analysis. 
%
The model consists of two cylinders of length $L$ and radius $a$ at distance
$d$. Finite size effects are reduced by having the two cylinders inside a fluid
sphere of radius $100L$. No-slip boundary conditions are set on the cylinder
surfaces.  We then calculate Stokes drag on the cylinders in the case of rigid
and relative motion at different $d$ values. Hydrodynamic coupling is then
obtained as half the difference between rigid and relative mobilities.  The
resulting values for coupling $G^{xx}$ are reported as solid circles in Fig.
\ref{G} showing a remarkable good agreement with the analytical expression in
(\ref{Gth}) also plotted as a solid line.
%

\begin{figure}[ht]
\includegraphics[width=.45\textwidth]{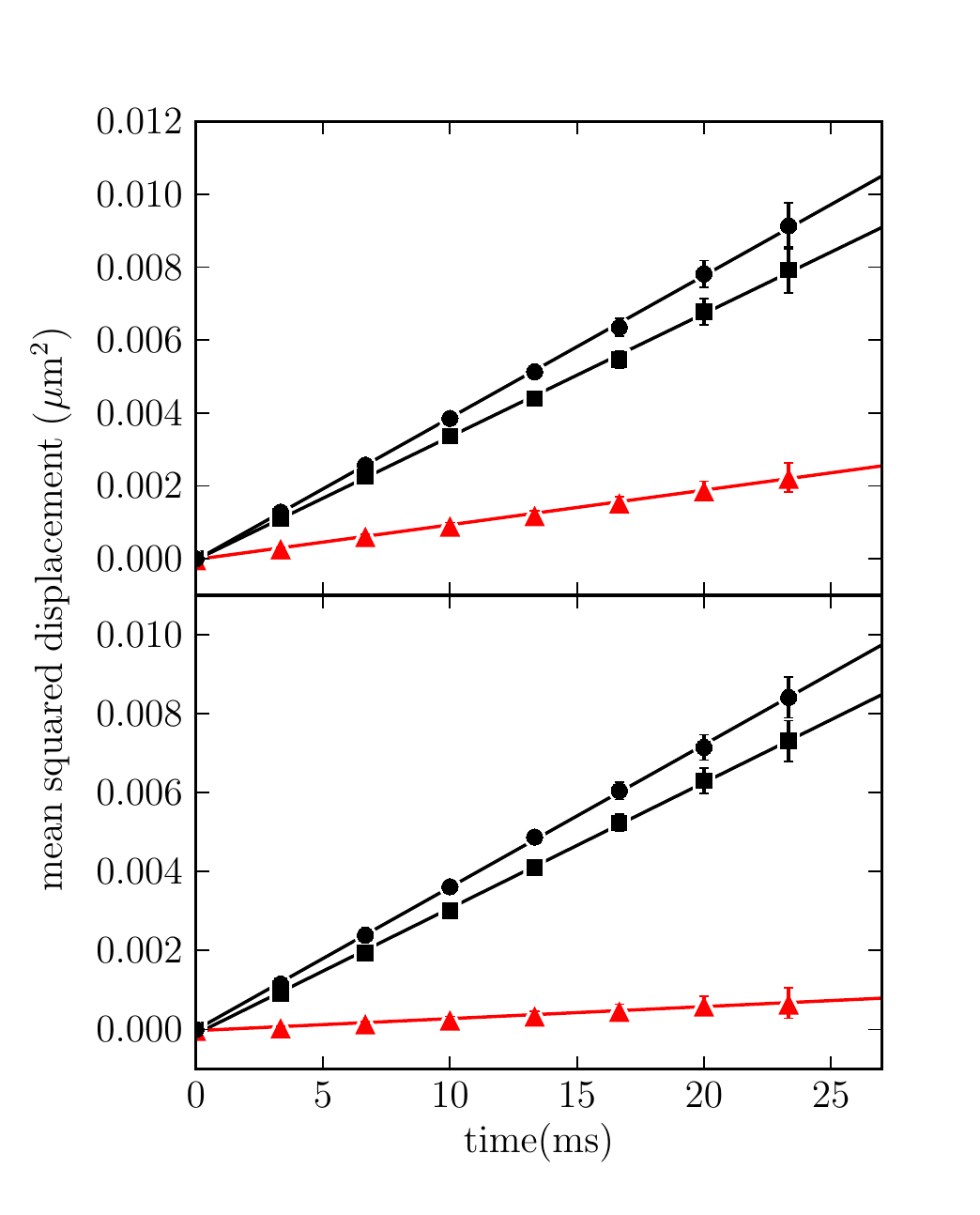}
\caption{Mean squared displacements of freely diffusing rods starting off a
parallel aligned configuration generated with holographic optical tweezers.
Top and bottom panel refer respectively to interparticle distances of
3.5 and 20 $\mu$m. Solid circles and squares represent the diagonal terms of the generalized MSD while solid triangles refer to the off diagonal terms. Linear fits are also reported as solid lines.
}
\label{msd}
\end{figure}

The chance of observing two rods, occasionally aligned in parallel at some
given distance, clearly makes the problem practically impossible to approach
experimentally by simple video microscopy observations.  However holographic tweezers
provide an ideal tool to directly verify our results by trapping and orienting two
slender objects in well defined and reproducible relative configurations. We
used holographic tweezers to manipulate silica microrods of about 10 $\mu$m in
length and 150 nm in radius \cite{carberry}.  Multiple optical traps are obtained by  shaping
an expanded  CW laser beam (532 nm) with a spatial light modulator (Holoeye
LC-R 2500). The shaped beam is then focused by a microscope objective of high
numerical aperture (Nikon Plan Apo 100x, NA 1.4) onto the target array of
trapping spots. Optimal intensity distribution are generated by GSW algorithm
running on an NVIDIA GPU (GTX 480) \cite{gsw, cuda}.  Each rod is held in two
traps independently and dynamically reconfigurable (Fig. \ref{rods}b).  The two rods can be
aligned and moved at different distances.  We limit ourselves to the computed
case of center of mass dynamics of parallel aligned rods.  Using a chopper on
the trapping beam, we periodically release the rods every 1/15 s and
subsequently start recording bright field images at 300 fps for 1/30 s.  The
horizontal $x$ coordinate of each rod is tracked by fitting the row average
intensity profile of cropped rods images.  A straightforward way to obtain
hydrodynamic couplings is that of tracking rods free Brownian motion. We have
estimated and experimentally verified that our rods have a rotational diffusion
coefficient of about 0.02 rad$^2$/s. That means that the mean squared angular
displacement during acquisition time is about $\pi$/100. It is therefore safe
to neglect rotational-translational coupling \cite{doi, yodh} and project on
the $x$ axis the mobility problem  with stochastic forces:
\begin{equation}
\begin{array}{c}
\left(
\begin{array}{c}
\dot x_1(t)\\ \dot x_2(t)
\end{array}
\right)
=\mathds M^{xx}\cdot
\left(
\begin{array}{c}
\eta_1(t)\\ \eta_2(t)
\end{array}
\right)\\[5mm]
\mathds M^{xx}=
\left(
\begin{array}{cc}
m_1 & G^{xx}(d)\\
G^{xx}(d) & m_2 
\end{array}
\right)
\end{array}
\end{equation}
where $m_1$ and $m_2$ are the transverse mobilities of the two rods while
$\eta_1$ and $\eta_2$ are stochastic forces with zero average and correlation:
\begin{equation}
\langle \eta_i(0)\eta_j(t)\rangle=2 k_B T (\mathds M^{xx})^{-1}_{ij}\; \delta(t)
\end{equation}
This is a coupled Langevin equation leading to a displacements covariance
matrix \cite{murphy}:
\begin{equation}
\begin{array}{c}
\langle (x_i(t)-x_i(0))(x_j(t)-x_j(0))\rangle=2 D_{ij} t\\[5mm]
D_{ij}=k_B T\; \mathds M^{xx}_{ij}
\label{D}
\end{array}
\end{equation}
Diagonal terms represent the mean squared displacements of the two rods.  The
corresponding diffusion coefficients provide a direct measurement of single
particle mobilities. On the other hand hydrodynamic couplings are easily
extracted from the diffusion coefficients of the off diagonal terms.  In Fig. \ref{msd} we report the two single particle mean squared displacements
toghether with the crossed term for interparticle distances of 3.5 and 20
microns. Diffusion coefficients can be easily extracted by fitting with a
straight line. Single particle diffusion coefficients don't show a systematic
dependence on interparticle distance and their deviations are mainly
attributable to small differences in rod lengths.  Microrods in our sample have
an average length of 10.5 $\mu$m with a standard deviation of 2.5\%.  Tracking
single rod Brownian motion, we extract an average transverse mobility of 42
$\mu$m/s pN. By theoretical predictions, this corresponds to an average rod
thickness of about 200 nm, which is well compatible with the nominal pore size
(300 nm) used for rod's growth \cite{carberry}.
Turning now to hydrodynamic couplings, the diffusion coefficient of crossed
terms  directly provides a measure of $G^{xx}(d)$ through (\ref{D}). Indeed we
found an excellent agreement with theoretical and numerical predictions as
shown in Fig.  \ref{G}. It is worth to note that coupling values are expected
to be much less affected by the actual rod thickness than the single rod
mobilities, as can be understood noticing that the thickness never entered in
the arguments leading to (\ref{Gth}).  In particular, the results in Fig.
\ref{G} are expected to remain valid even in the case of vanishing rods
thickness, as could be the case for single walled carbon nanotubes,
microtubules or short straight DNA segments.

In conclusion, we have directly measured hydrodynamic interactions between
freely diffusing micro-rods. We demonstrated by direct experimental observation
and numerical finite element analysis that two parallel aligned microrods
interact like point particles in 3D when they're much farther then their
lentgh, and like point particles in 2D for short distances. We also derived a
simple analytical expression that reproduces very well both
experimental and numerical data. Although the validity of our results is
expected to hold even for much thinner, and interesting objects, like nanotubes
or microtubules, an experimental investigation of hydrodynamic coupling in
those systems remains an open and challenging problem.
This work was partially
supported by IIT-SEED BACTMOBIL project and MIUR-FIRB project RBFR08WDBE.
\bibliographystyle{myapsrev}

\begin{thebibliography}{20}

\bibitem{happel}
J. Happel and H. Brenner, Low Reynolds Number Hydrodynamics, 
Kluwer Academic, Dordrecht, (1983).

\bibitem{lighthill}
J. Lighthill,  Mathematical Biofluiddynamics, SIAM, Philadelphia (1975). 

\bibitem{reich}
M. Reichert and H. Stark, Eur. Phys. J. E {\bf 17}, 493 (2005).  

\bibitem{polin}
M. Polin, I. Tuval, K. Drescher, J.P. Gollub, and R. E. Goldstein,
Science {\bf 325}, 487 (2009).

\bibitem{crocker}
J. Crocker, J Chem Phys, {\bf 106}, 2837 (1997).

\bibitem{meiners}
J. Meiners and S. Quake, Phys. Rev. Lett., {\bf 82}, 2211 (1999).

\bibitem{ring}
R. Di Leonardo, et al.,
Phys. Rev. E, {\bf 76}, 258301 (2007). 


\bibitem{tan}
S. Tan, H.A. Lopez, C.W. Cai, Y. Zhang,
Nano Lett. {\bf 4}, 1415 (2004)

\bibitem{onofrio}
O.M. Marag\`o, et al.
Nano Lett., {\bf 8}, 3211 (2008).

\bibitem{marago}
O.M. Marag\`o et al.
ACS Nano, {\bf 4},  7515 (2010).

\bibitem{plewa}
J. Plewa, E. Tanner, D. Mueth, D. Grier,
Opt. Express {\bf 12}, 1978 (2004).

\bibitem{grier}
R. Agarwal, K. Ladavac, Y. Roichman, G. H. Yu, C. M. Lieber, and D. G. Grier,
Opt. Express {\bf 13}, 8906 (2005).

\bibitem{simpson}
S.H. Simpson and S.Hanna, JOSA A, {\bf 27}, 1255 (2010).

\bibitem{ikin}
L. Ikin, D.M. Carberry, G.M. Gibson, M.J. Padgett, and M.J. Miles,
New J. Phys. {\bf 11}, 023012 (2009).

\bibitem{carberry}
D.M. Carberry et al., Nanotechnology, {\bf 21}, 175501, (2010).

\bibitem{Kim}
S. Kim and S. J. Karrila, Microhydrodynamics: princi- 
ples and selected applications, Dover (2005). 

\bibitem{2d}
R. Di Leonardo, S. Keen, F. Ianni, J. Leach, M. Padgett, Ruocco G.,
Phys. Rev. E, {\bf 78}, 031406, (2008). 

\bibitem{gsw}
R. Di Leonardo, F. Ianni, G. Ruocco,
Optics Express, {\bf 15}, 1913, (2007).

\bibitem{cuda}
S. Bianchi, R. Di Leonardo,
Comp. Phys. Comm., {\bf 181}, 1444, (2010). 
%
\bibitem{doi}
M. Doi and S.F. Edwards, The theory of polymer dynamics, Oxford University
Press, New York (1986).
%
\bibitem{yodh}
Y. Han, A.M. Alsayed, M. Nobili, J. Zhang, T.C. Lubensky, A.G. Yodh,
Science, {\bf 314}, 626 (2006).
%
\bibitem{murphy}
T.J. Murphy and J.L. Aguirre, J. Chem. Phys. {\bf 57}, 2098 (1972).

\end{thebibliography}

\end{document}